\documentclass[12pt]{iopart}
\usepackage{iopams}  
\usepackage{hyperref}
\usepackage{color}
\usepackage[a4paper]{anysize}

\definecolor{darkred}{rgb}{0.5,0.0,0.0}
\definecolor{darkgreen}{rgb}{0.0,0.5,0.0}
\definecolor{darkblue}{rgb}{0.0,0.0,0.5}
\hypersetup{colorlinks=true,
citecolor=darkred,
linkcolor=darkgreen,
urlcolor=darkblue}
\marginsize{3cm}{2.5cm}{2.0cm}{4.0cm}

\usepackage{graphicx}
\usepackage{bm}
\usepackage[applemac]{inputenc}

\newcommand{\re}{\ensuremath{\mathrm{Re}\,}}
\newcommand{\im}{\ensuremath{\mathrm{Im}\,}}
 
\begin{document}   

\title{Optical conductivity, Drude weight and plasmons in twisted graphene bilayers}
\author{Tobias Stauber$^1$, Pablo San-Jose$^2$, Luis Brey$^2$}
\address{$^1$Departamento de F\'{\i}sica de la Materia Condensada, INC and IFIMAC, Universidad Aut\'onoma de Madrid, E-28049 Madrid, Spain\\$^2$Instituto de Ciencia de Materiales de Madrid (ICMM-CSIC), Cantoblanco, 28049 Madrid, Spain}
%\affiliation{$^1$Departamento de F\'{\i}sica de la Materia Condensada, INC and IFIMAC, Universidad Aut\'onoma de Madrid, E-28049 Madrid, Spain\\$^2$Instituto de Ciencia de Materiales de Madrid (ICMM-CSIC), Cantoblanco, 28049 Madrid, Spain}

\date{\today} 

\begin{abstract}
We numerically calculate the optical conductivity of twisted graphene bilayers within the continuum model. To obtain the imaginary part, we employ the regularized Kramers-Kronig relation allowing us to discuss arbitrary twist angles, chemical potential and temperature. We find that the Drude weight $D$ as function of the chemical potential $\mu$ closely follows the shell structure of twisted bilayer displayed by the density of states. For certain angles, this results in a transport gap $D=0$ at finite $\mu$. We also discuss the loss function which, for low doping, is characterized by acoustic interband "plasmons" and transitions close to the van Hove singularities. For larger doping, the plasmon mode of decoupled graphene bilayer is recovered that is damped especially for small wave numbers.
\end{abstract}

\maketitle

\section{Introduction}

Since the realization that neutral graphene monolayers display a universal optical absorption of $\pi \alpha\approx 2.3\%$ in the infrared and visible spectrum (with $\alpha\approx 1/137$ the fine structure constant),\cite{Nair:S08,Mak08,FalkovskyPer07,Stauber:PRB08a} interest in the interactions between light and graphene has grown rapidly.\cite{Avouris10,Bonaccorso10,Koppens11} A remarkable aspect of this absorption, other than its suggestive relation to the universal constant $\alpha$, is its high value, given the ultimate thinness of a graphene monolayer. In fact, it corresponds to an attenuation factor that is hundreds of times larger than that of good metals, such as gold and can be even enhanced up to 10\% for photon energies around 4.5eV due to transitions close to the van Hove singularity.\cite{Yang09,Kravets10}

The unique feature of neutral graphene responsible for its strong optical response is the existence of low energy particle-hole interband excitations with zero momentum, by virtue of the gapless Dirac spectrum, that act as an effective sink for photons of arbitrarily low energy. By the same token, all thermally excited plasmons in undoped graphene are damped by this excitation density, but nevertheless remain well-defined.\cite{Vafek06} The unique bipolar tunability of graphene further allows one to adjust these optical properties quite simply, by shifting the Fermi energy $\mu$ away from the Dirac point, be it electrostatically or through chemical doping. This opens a gap of energy $2\mu$ in the particle-hole excitations, making graphene transparent below this energy, useful, e.g., for broadband optical modulation.\cite{Liu11} It also allows for the emergence of undamped plasmons.\cite{Wunsch06,Hwang07,Ju12,Chen12,Fei12}

The spectral features involved in the unusual optical properties outlined above are also present in graphene bilayers.\cite{Abergel07} These are often classified according to their stacking, which in the absence of strain have three possible (meta) stable configurations: AB'-type (or BA', also known as Bernal stacking), AA'-type, and twisted. The former two are minimal stackings (unit cell with only four carbon atoms) in which half or all sites in opposite layers are vertically aligned, respectively. The twisted bilayer is the most generic type, and is defined by an relative rotation of angle $\theta$ between the two layers. 
%This produces a Moiré pattern of local stacking that alternates between AB', BA' and AA. 
Each of the three stacking types exhibits a different low-energy electronic structure, although all three react strongly to light like the monolayer. The twisted bilayer, however, and in particular the case of low twist angles (around and below $1º$), has the richest optical response, a consequence of its complex low energy spectral properties. The response, moreover, is highly tuneable through the angle $\theta$, an additional control parameter not available in the other graphenes. % induced by the Moiré pattern. 
As a striking example, the strong frequency-dependence of twisted bilayer conductivity and thus reflectivity leads to a visible coloration under white light, that varies with angle $\theta$, of flakes on 100 nm thick $\mathrm{SiO}_2$, as recently observed experimentally.\cite{Campos-Delgado:13}

In this work we study the optical response of twisted graphene bilayers at different twist angles and temperature responsible for this effect. We analyze its optical conductivity at and away from neutrality, the Drude weight and plasmon spectrum. Some of these aspects, specifically the real part of the conductivity, have been studied in recent works for large-angle twisted bilayers.\cite{Moon:13,Tabert:13} We will make connection with these, and extend them into the low angle regime.

The paper is organized as follows. In Section II, we introduce the continuum model of twisted graphene bilayer and discuss the electronic structure. In Section III, we introduce the Kubo formalism and also comment on the regularization of the Kramers-Kronig relation for the conductivity, necessary in the case of unbounded Dirac Fermions. In Section IV and V, we present our results for the optical conductivity and the related Drude weight for different angles, chemical potential and temperature. In Sec. VI, we discuss the plasmonic spectrum based on the local conductivity by plotting the energy loss function and close with conclusions.

\section{Continuum model of a twisted graphene bilayer}

Twisted graphene bilayers are a peculiar structure from a crystallographic point of view, since for a generic (non-commensurate) angle the lattice is not exactly periodic, although crystalline order within each layer is extraordinarily robust. This is a consequence of the highly anisotropic binding forces in these systems, also characteristic of graphite. In principle only commensurate twist angles $\theta=\theta_{mn}$, where \cite{Santos:PRB12}
\begin{equation}
\label{conmensurable}
\mathrm{cos}(\theta_{mn})=\frac{3m^2+3 m n+n^2/2}{3m^2+3 m n+n^2}
\end{equation}
for integer $m,n$, are amenable to band structure calculations by application of the Bloch theorem. It turns out, however, that the bandstructure at low energies depends continuously on the value of the angle $\theta_{n,m}$ itself, not on the specific $n$ and $m$ (note that the set of possible commensurate angles $\theta_{mn}$ is dense). This is a consequence of the fact that the physical interlayer distance $d>0.3$ nm is greater than the intralayer carbon-carbon distance $a_{cc}=0.14$ nm. In this regime, all properties associated to precise commensurability are efficiently averaged out using microscopic tight-binding models. Consequently, a simple, low energy continuum theory can be formulated that may be applied to an arbitrary twist angle $\theta$, and obviates all crystallographic complications to a very good approximation.\cite{Mele:PRB10,Santos:PRB12}

The first ingredient in the continuum model is the momentum shift between Dirac cones in different layers due to the rotation, $\Delta K=2|K|\sin(\theta/2)$, where $|K|=4\pi/3a_0$ and 
$a_0=\sqrt{3}a_{cc}$ is the monolayer Bravais period. The second non-trivial ingredient is the smooth spatial modulation in the interlayer coupling. The twist $\theta$ produces a triangular Moiré pattern in the local stacking, of period $L_M=a_0/[2\sin(\theta/2)]$. Hence, the coupling between layers becomes position dependent, alternating periodically in space between AB', BA' and AA'. The modulation is smooth on the scale $a_0$, so the two valleys in each layer remain decoupled in the twisted bilayer. For each valley, 
the continuum model Hamiltonian reads \cite{Santos:PRL07,Santos:PRB12,San-Jose:PRL12}
\[
H=\left(\begin{array}{cccc}
0 & \hbar v_F \Pi_+^\dagger & V^*(\bm{r}) & V^*(\bm{r}-\bm{\delta r}) \\
\hbar v_F\Pi_+ & 0 & V^*(\bm{r}+\bm{\delta r}) & V^*(\bm{r})    \\
V(\bm{r}) & V(\bm{r}+\bm{\delta r}) & 0 & \hbar v_F \Pi_-^\dagger\\
V(\bm{r}-\bm{\delta r}) & V(\bm{r}) & \hbar v_F\Pi_- & 0   \\
\end{array}\right)
\]
This matrix is written in the local basis $A, B, A', B'$, comprised of the two sites in the unit cell of each of the two layers. Here, the Fermi velocity is $v_F=(\sqrt{3}/2)a_0\gamma_0/\hbar\approx 9\cdot 10^5$ m/s, and $\Pi_\pm=\left[k_x+i(k_y\mp\Delta K/2)\right]e^{i\pm\theta/2}$. The function $V(\bm{r})=\frac{1}{3}\gamma_\perp\sum_{i=1}^3 e^{i\bm{g}_i\cdot\bm{r}}$ describes the periodic spatial variation of the interlayer coupling, with $\gamma_\perp\approx 0.33 \mathrm{eV}$. 
The Moiré lattice vectors are denoted by $\bm{a}_{1,2}$ (so $L_M=|\bm{a}_{1,2}|$), and conjugate momenta are $\bm{g}_{1,2}$, such that $\bm{g}_i\cdot\bm{a}_j=2\pi \delta_{ij}$ (we also define $\bm{g}_3=0$). The AA' sublattice is centered at the origin, with the AB'/BA' sublattices offset given by $\pm\bm{\delta r}=(\bm{a}_1-\bm{a}_2)/3$. 

We first review the bandstructure phenomenology of the twisted bilayer. The continuum model has two distinct regimes, denoted as large and small angles, each with very different phenomenology. The large angle regime, relevant for $\theta$ greater than a few degrees, is characterised by two low energy Dirac cones per valley at the superstructure $K$ and $K'$, with a reduced Fermi velocity  \cite{Castro:PRL07,Suarez-Morell:PRB10,Bistritzer:P11}
\begin{equation}
v_F^*\approx v_F[1-9\alpha^2+\mathcal{O}(\alpha^4)],
\label{veff}
\end{equation}
where $\alpha=\gamma_\perp/6\epsilon_M$ is an expansion parameter that represents the dimensionless interlayer coupling, and $\epsilon_M(\theta)=\hbar v_F\Delta K/2=(2\pi/3)\hbar v_F/L_M$ is the energy scale associated to the Moiré period $L_M$. Also, a low energy van-Hove singularity at energy $\epsilon_{vH}\approx\epsilon_M-\gamma_\perp/3$ is created \cite{Santos:PRL07,Li:NP10,Luican:PRL11,Yan:PRL12, Brihuega:12}.

At an angle $\theta_1^m\approx 1.05º$, known as the ``first magic angle", $v_F^*$ becomes zero, the van-Hove singularity degenerates into a narrow miniband around zero energy. This is the onset of the small angle regime. 
From this point on, the electronic structure becomes considerably complex. The Fermi velocity becomes finite again, as the miniband moves away from zero energy. Higher energy van-Hove singularities appear and approach zero at subsequent magic angles $\theta_{2,3,4\dots}^m$, where the Fermi energy vanishes once more. 
%In the intervals between the first five magic angles ($1.05º>\theta>\approx 0.23º$), Dirac cones reappear

\begin{figure}
   \centering
   \includegraphics[width=0.8\columnwidth]{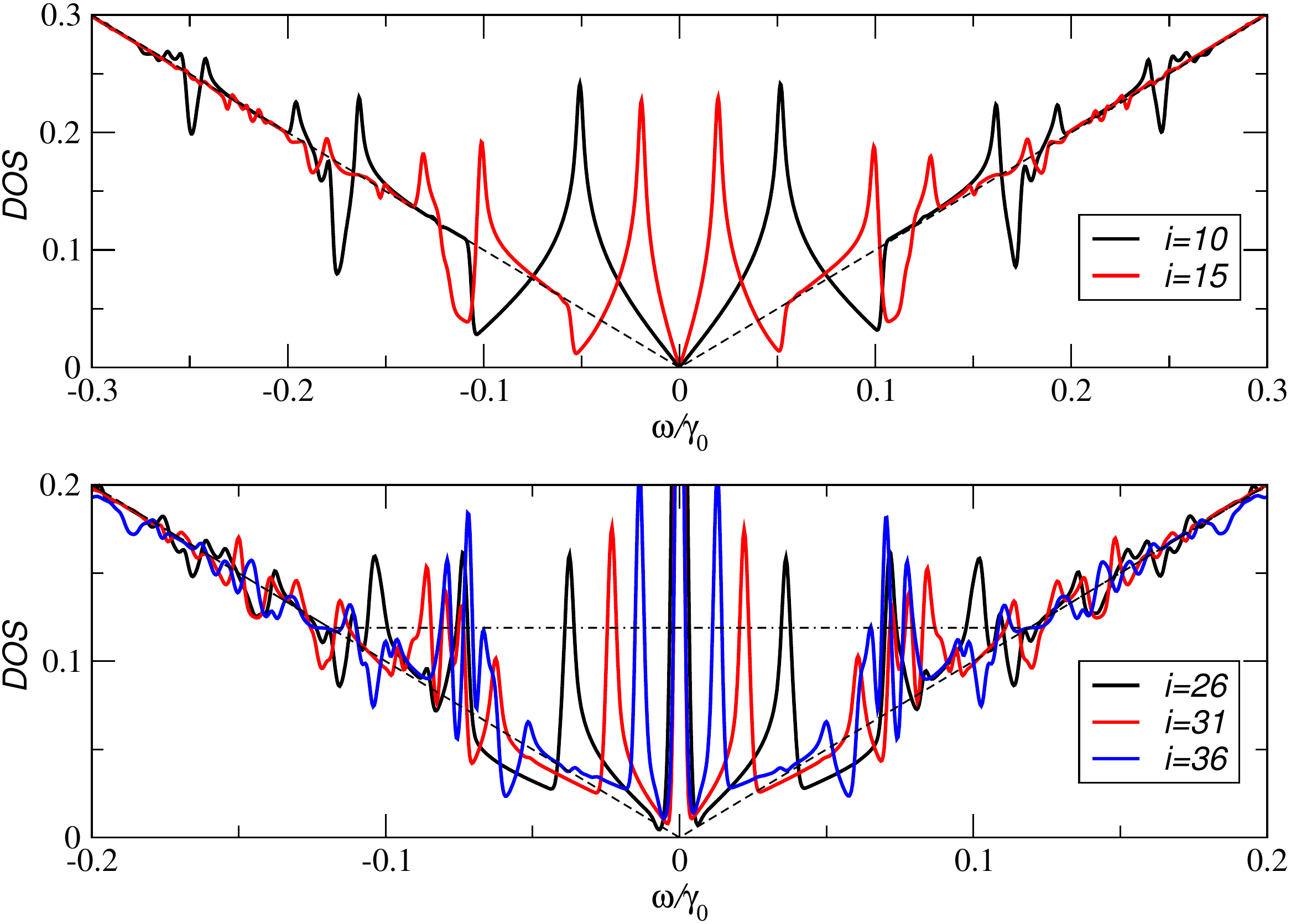} 
   \caption{color online: Density of States (DOS) for different angles $\theta_i$ in units of $\rho_0=\frac{4\gamma_0}{\pi (\hbar v_F)^2}$. Upper panel: $\theta=3.15º$ (black) and $\theta=2.13º$ (red). Lower panel: $\theta=1.25º$ (black), $\theta=1.05º$ (red) and $\theta=0.91º$ (blue). The dashed and dotted-dashed line resemble the DOS of decoupled bilayer and AA-stacked bilayer, respectively. The index $i$ is related to the angle by Eq. (\ref{anglei}).}
   \label{fig:dos}
\end{figure}

In Fig. \ref{fig:dos}, the density of states (DOS) is shown for various angles labeled by the index $i$
\begin{equation}
\label{anglei}
\cos(\theta_i)=1-\frac{1}{2(3i^2+3i+1)}\;,
\end{equation}
obtained from Eq. (\ref{conmensurable}) with $m=n+1=i$. The curves are normalized by $\rho_0=\frac{4\gamma_0}{\pi (\hbar v_F)^2}$  such that the DOS of two uncoupled graphene layers is just a straight line with slope 1 (dashed line). Note that the DOS is almost but not exactly symmetric for positive and negative energies as a consequence of the underlying (not electron-hole symmetric) Hamiltonian. 

In the upper panel, we show the DOS for "large" angles $\theta$ with $i=10$ (blue) and $i=15$ (red). Around the neutrality point, the spectrum is characterized by a van Hove singularity followed by a step singularity, forming a closed shell. For small energies the system is thus approximately described by a hexagonal tight-binding model with renormalized band-width. This shell structure is somehow repeated for larger energies until it fades away and the DOS for decoupled bilayer graphene is recovered. Note that the weight of the van Hove singularity is slightly decreasing with increasing (intermediate) $i$. 

For small angles around the first magic angle $\theta_1^m=1.05º$ ($i=31$, lower panel in Fig. \ref{fig:dos}), more distinct van Hove singularities appear and the height of the first van Hove singularity is strongly enhanced with increasing $i$. A close-up of this energy regime in shown on the left hand side of Fig. \ref{fig:DrudeDos}. 
%The dashed-dotted line resembles the DOS of AA-stacked graphene,\cite{Roldan13} which can be interpreted as the average of the DOS of twisted bilayer graphene over small angles.

\section{Linear response theory}
The above phenomenology in the electronic bandstructure has an associated signature in the optical response of the system as the angle decreases. The basic quantity governing the linear optical response of any electronic system is the optical conductivity $\sigma_{\alpha\beta}(\omega)$ (where $\alpha,\beta=x,y$), which gives the electric current arising in response to a weak and uniform, time-dependent electric field of frequency $\omega$, $\bm{E}(t)=\bm{E}(\omega)e^{i\omega t}$, 
\begin{equation}
j_\alpha(\omega)=\sum_\beta\sigma_{\alpha\beta}(\omega) E_\beta(\omega)\;.
\end{equation}
The optical conductivity may be written, using linear response theory, in terms of the current response $\chi_{j_\alpha j_\beta}(\bm{q},\omega)$ at zero-momentum,
\begin{equation}
\sigma_{\alpha\beta}(\omega)=i\frac{e^2}{\omega+i\delta}\chi_{j_\alpha j_\beta}(\bm{q}=0,\omega),
\label{sigma}
\end{equation}
where, as is conventional, $\delta$ denotes a positive infinitesimal. The current response function $\chi_{j_\alpha j_\beta}$ consists, in general, of two terms: the diamagnetic and the paramagnetic contribution.\cite{Giuliani:05} The general formalism for obtaining the diamagnetic contribution in a tight-binding model is outlined in Ref. \cite{Stauber10b}, however, in the continuum model for the twisted bilayer, as in the monolayer, the diamagnetic contribution vanishes due to the fact that $\partial^2 H/\partial k_\alpha^2=0$. Hence, we have simply the paramagnetic contribution, given by the Kubo formula,
\begin{eqnarray}
\chi_{j_\alpha j_\beta}(\bm{q},\omega)&=&g_s g_v\int\frac{d^2\bm{k}}{(2\pi)^2} \sum_{mn}
\frac{n_F(\epsilon_{m,\bm{k}})-n_F(\epsilon_{n,\bm{k}+\bm{q}})}{\hbar\omega-\epsilon_{n,\bm{k}+\bm{q}}+\epsilon_{m,\bm{k}}+i\delta}\nonumber\\
&\times& \langle m,\bm{k}|j_\alpha|n,\bm{k}+\bm{q}\rangle \langle n,\bm{k}+\bm{q}|j_\beta|m,\bm{k}\rangle\;. 
\label{Kubo}
\end{eqnarray}
Here, $g_s=g_v=2$ are the spin and valley degeneracies. States $|m,\bm{k}\rangle$  are eigenstates of $H$ in subband $m$ and of momentum $\bm{k}$ in the first Brillouin zone of the superstructure. Their eigenenergies are $\epsilon_{m\bm{k}}$ and $n_F$ is the Fermi function. The paramagnetic current operator is $j_\alpha=-\partial H/\partial k_\alpha$. In the continuum model with local interlayer coupling, this current does not mix layers and thus independent of the coupling $\gamma_\perp$. For a system with time-reversal symmetry, the conductivity tensor is diagonal and due to the threefold rotational symmetry of (twisted) bilayer graphene proportional to the unity tensor. We will, therefore, drop the tensor indices in what follows.

The real part of the conductivity $\sigma(\omega)$ represents the power dissipation of the system upon incidence of light at frequency $\omega$, while its imaginary part represents inductive retardation of the current with respect to the driving field. From Eq. (\ref{sigma}), it is obvious that the conductivity is ill-defined at $\omega=0$ and $\delta\rightarrow0$ if $\chi_{jj}({\bm q}=0,\omega=0)\neq0$.  It is thus customary to split up the conductivity into a regular part and a term containing the delta-singularity:
\begin{eqnarray}
\sigma(\omega)&=&\pi D\delta(\omega)+\sigma_\mathrm{reg}(\omega)\\
\sigma_\mathrm{reg}(\omega)&=&\frac{ie^2}{\omega}\chi_{jj}(\bm{q}=0,\omega)
\end{eqnarray}
with the Drude weight (or charge stiffness) $D$ defined as
\begin{equation}
\label{Drude}
D=e^2 \lim_{\omega \to0}\re\chi_{jj}(\bm{q}=0,\omega)=\lim_{\omega\to 0} \omega \,\im\sigma(\omega)\;.
\end{equation}
Above we used the fact that the imaginary part of the current response is an odd function due to causality. We thus have $\im\sigma=\im\sigma_\mathrm{reg}$ and will drop the subindex when confusion is unlikely.

\subsection{Graphene monolayer}
In the case of a graphene monolayer, described within a continuum Dirac model, we have the famous zero temperature result ($\mu$ is the chemical potential with respect to the neutrality point)
\begin{eqnarray}
\label{Resigma0}
\re \sigma_\mathrm{reg}(\gamma_0\gg\omega>2\mu)&=&\sigma_0=\frac{\pi}{2}\frac{e^2}{h}\\
\re \sigma_\mathrm{reg}(\omega<2\mu)&=&0\nonumber.
\end{eqnarray}
This universal value $\sigma_0$ for the Dirac gas includes valley and spin degeneracy factors, and is valid for energies $\hbar \omega\sim1$eV well below the $\gamma_0\sim 3$eV van-Hove singularity of the monolayer. As $\omega$ approaches this scale, a  discrete model must be employed to compute the monolayer $\sigma_{0}$ (see Ref. \cite{Stauber:PRB08a} for general results). 

The imaginary part of the conductivity, in the same limit of low frequencies, reduces to\cite{Stauber:PRB08a,FalkovskyPer07}
\begin{equation}
%\label{Imsigma0}
\im\sigma(\omega\ll \gamma_0)=\frac{4}{\pi}\frac{\mu}{\hbar\omega}\sigma_0+\frac{1}{\pi}\log\left|\frac{\hbar\omega-2\mu}{\hbar\omega+2\mu}\right|\sigma_0\;,
\end{equation}
where $\mu$ is the Fermi energy, measured from the Dirac point (charge neutrality). Using Eq. (\ref{Drude}), we have a Drude weight 
\begin{equation}
\label{DrudeDirac}
D_0=\frac{4\mu}{\pi\hbar}\sigma_0\;
\end{equation}
which vanishes at neutrality. Note that the above expression only scales with the chemical potential and no other material constant (Fermi velocity) is involved. The Drude weight for twisted bilayer as function of $\mu$ will thus be independent of the twist angle in the regime where the electronic spectrum can be described by a (renormalized) Dirac model. This breaks down close to the first magic twist angle with the emergence of a flat band.

\subsection{Twisted bilayer}

In the more general case of a twisted bilayer, the computation of $\sigma(\omega)$ becomes a numerical task and it is usually simpler to compute $\re\sigma_\mathrm{reg}$ than $\im \sigma$, since, by virtue of Eq. (\ref{Kubo}), the former reduces to an integral constrained to eigenstates within $\hbar\omega$ of the Fermi energy
\begin{eqnarray}
\re \sigma_\mathrm{reg}(\omega)&=&\frac{16\pi\sigma_0}{\omega}\int \frac{d^2\bm{k}}{(2\pi)^2} \sum_{mn}\left[n_F(\epsilon_{n,\bm{k}})-n_F(\epsilon_{m,\bm{k}})\right]\nonumber\\ 
&\times& \left|\langle m,\bm{k}|j_x|n,\bm{k}\rangle\right|^2
\delta\left[\omega-(\epsilon_{m,\bm{k}}-\epsilon_{n,\bm{k}})/\hbar\right]\;.\nonumber
\end{eqnarray}
The real part of the conductivity differs from that of two decoupled monolayers only at frequencies smaller than or comparable to the interlayer coupling $\omega<\gamma_\perp$. Hence we have $\re \sigma_\mathrm{reg}(\omega\gg \gamma_\perp)\approx 2\sigma_0$. This result will be useful in the following discussion to obtain a regularized expression for the imaginary part of the conductivity.

To compute the imaginary part, and from that, the Drude weight $D$ through Eq. (\ref{Drude}), one could in principle employ Eq. (\ref{Kubo}). This is problematic,  since it involves a summation, even for small $\omega$, over the whole band, which in the continuum model extends to infinity. We will, therefore, make use of the the Kramers-Kronig (KK) relation, applicable to any response function. In terms of the conductivity, this yields
\begin{equation}
\im\sigma(\omega)=\frac{2}{\pi\omega}\mathcal{P}\int_0^\infty d\nu \frac{\nu^2\,\re\sigma_\mathrm{reg}(\nu)}{\omega^2-\nu^2}\label{KKIm}
%\re\sigma^\mathrm{reg}_{xx}(\omega)&=&\frac{2}{\pi}\mathcal{P}\int_0^\infty d\nu \frac{\nu\,\im\sigma_{xx}(\nu)}{\nu^2-\omega^2},
\end{equation}
where $\mathcal{P}$ denotes the Cauchy principal value. There is apparently no advantage to using Eq. (\ref{KKIm}) as compared to Eq. (\ref{Kubo}), since the integral above extends to infinity, too. Moreover, the integral is ill defined, since at high frequencies the continuum model yields a constant $\re \sigma_\mathrm{reg}(\omega\gg \gamma_\perp)\approx 2\sigma_0$.

We will thus define the imaginary part relative to this divergent background following the usual procedure of ultraviolet field theories. In fact, this is one way of obtaining Eq. (\ref{Imsigma0}) and was discussed more generally in the context of the f-sum rule in Ref. \cite{Sabio08}. The low-energy properties are not modified by this regularization and we can simply subtract the ground-state expectation value of the two independent graphene layers at zero doping. Since the imaginary part for the neutral system is zero for the continuum model, the final regularized definition of $\im\sigma$ simply reads
\begin{equation*}
\label{KKRegularization}
\im\sigma(\omega)=\frac{2}{\pi\omega}\int_0^\infty d\nu\frac{\nu^2(\re\sigma_\mathrm{reg}(\nu)-2\sigma_0)}{\omega^2-\nu^2}\;.
\end{equation*}

Numerically, Eq. (\ref{KKRegularization}) can be evaluated by introducing a finite cutoff $\Lambda$ for which $\re\sigma^\mathrm{reg}(\Lambda)\approx2\sigma_0$. Integrating the second term involving the "ground-state value" $2\sigma_0$ then yields the final formula:
\begin{eqnarray}
\im\sigma(\omega)&=&\frac{2}{\pi\omega}\left(\mathcal{P}\int_0^\Lambda d\nu \frac{\nu^2\re\sigma_\mathrm{reg}(\nu)}{\omega^2-\nu^2}+2\sigma_0\Lambda\right)\nonumber\\
&&+\frac{2\sigma_0}{\pi}\log\left(\frac{\Lambda-\omega}{\Lambda+\omega}\right)
\label{Imsigma0}
\end{eqnarray}
Note that this expression is independent of $\Lambda$ as it was obtained from Eq. (\ref{KKRegularization}). The extra terms compared to the original KK-relation (\ref{KKIm}) can be interpreted as the diamagnetic contribution that naturally arises in the Dirac model with finite band cutoff. 

Using this expression, the Drude weight $D$ in Eq. (\ref{Drude}) may be written as
\begin{equation}
\label{DrudeKK}
D=\frac{2}{\pi}\left(2\sigma_0\Lambda-\int_0^\Lambda d\nu \re\sigma_\mathrm{reg}(\nu)\right)
\end{equation}
In terms of $D$, Eq. (\ref{Imsigma0}) is then finally rewritten as
\begin{equation}
\im\sigma(\omega)=\frac{D}{\omega}+\frac{2\omega}{\pi}\mathcal{P}\int_0^\infty d\nu\frac{\re \sigma_\mathrm{reg}(\nu)-2\sigma_0}{\omega^2-\nu^2}\;.
\label{Imsigma}
\end{equation}
From Eq. (\ref{DrudeKK}), we can also deduce the f-sum rule
\begin{equation}
\label{SumRule}
\int_0^\Lambda d\omega \re\sigma(\omega)=2\sigma_0\Lambda
\end{equation}
which is independent of the angle $\theta$, chemical potential $\mu$ or temperature $T$.

\section{Optical conductivity}
\begin{figure}
   \centering
   \includegraphics[width=0.8\columnwidth]{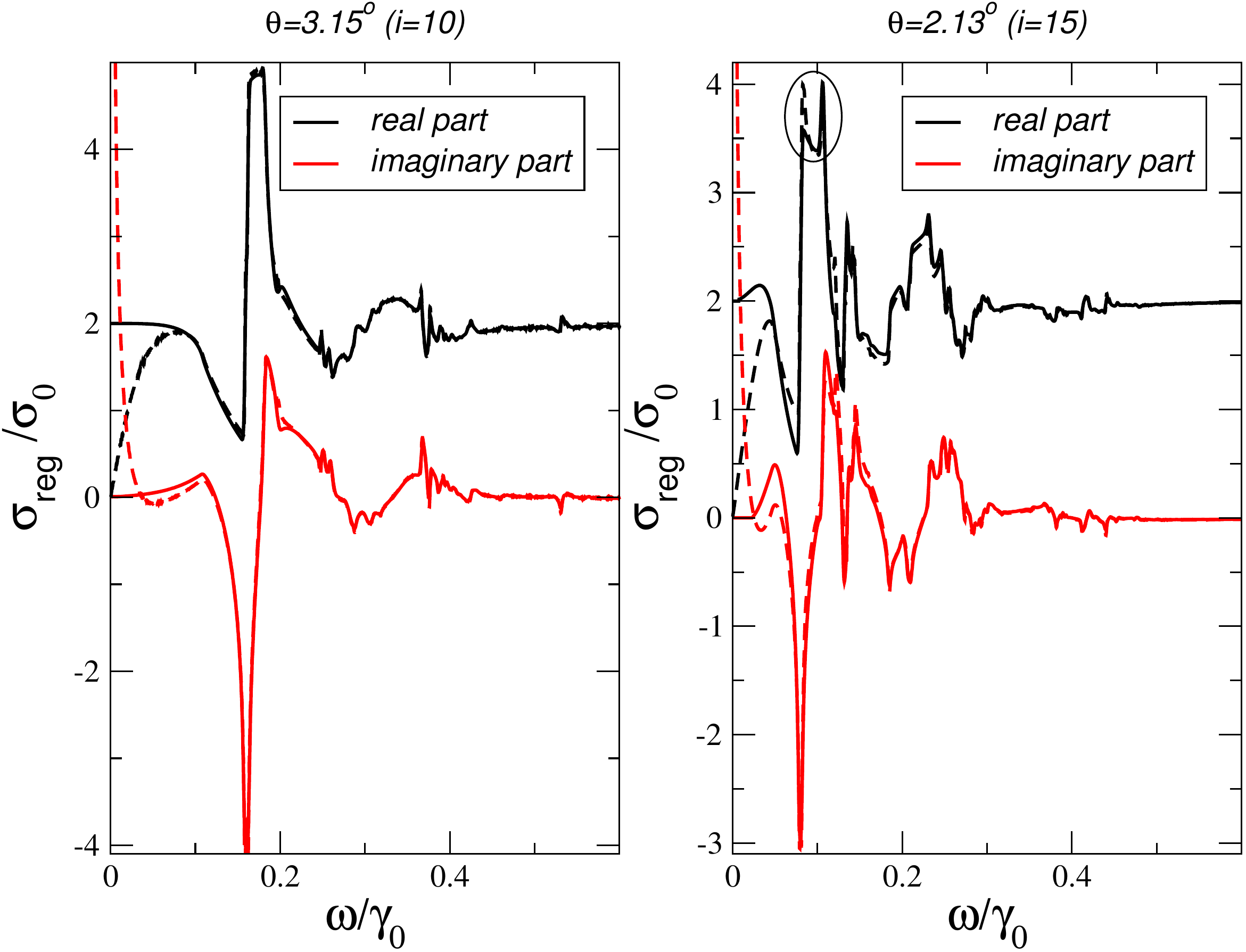} 
   \caption{(color online): Real (black) and imaginary (red) part of the optical conductivity of twisted bilayer graphene at neutrality $\mu=0$ for two different angles at $T=0$. Normalization constants are the universal monolayer conductivity $\sigma_0=\pi/2 e^2/h$, and $\gamma_0=2.78$ eV. Index $i$ is related to the angle by Eq. (\ref{anglei}). The dashed lines refer to finite temperature with $T=300K$.}
   \label{fig:sigma1}
\end{figure}

\begin{figure}
   \centering
    \includegraphics[width=0.8\columnwidth]{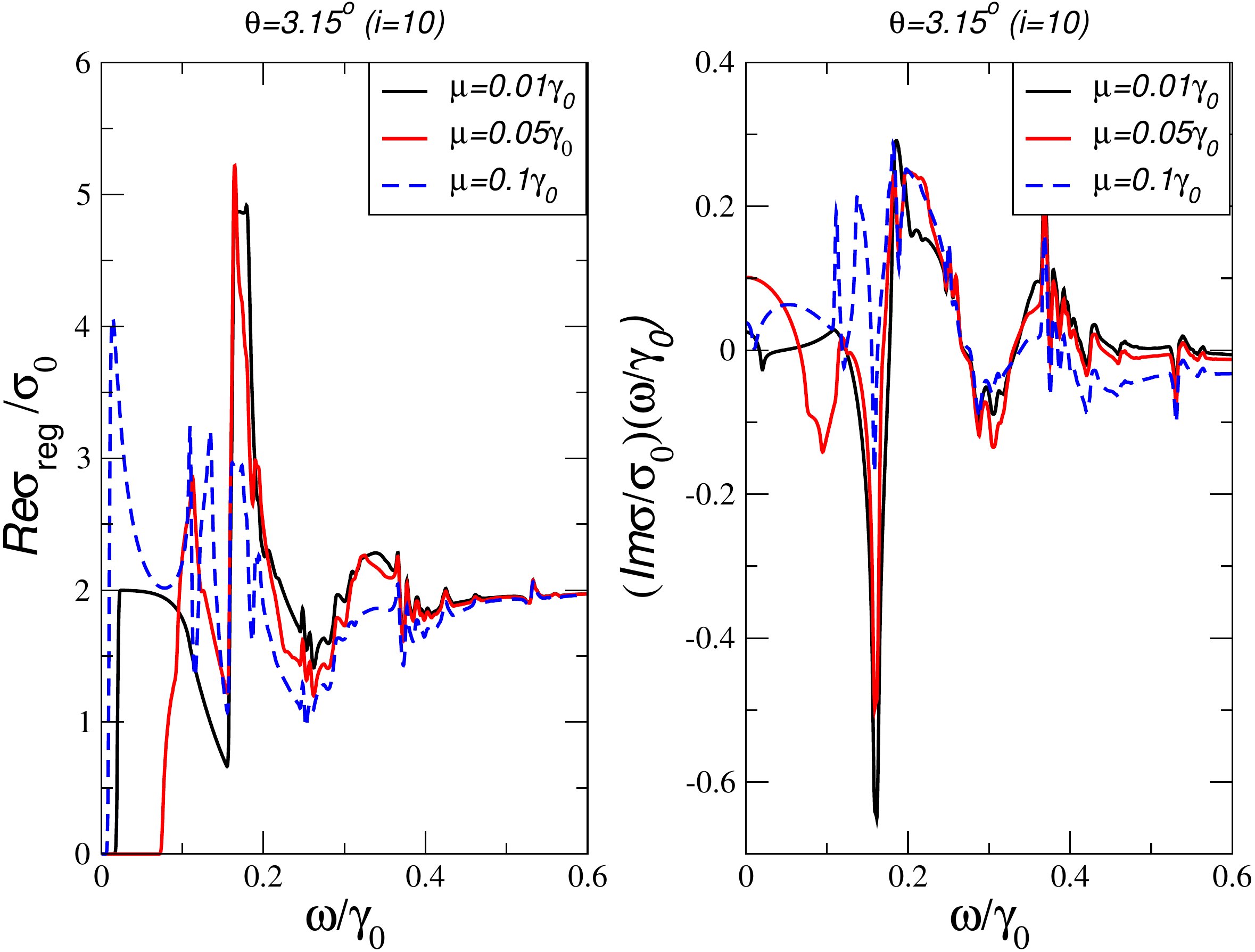} 
   \caption{(color online): Real (a) and imaginary (b) part of the optical conductivity at increasing value of the Fermi energy $\mu$, as measured from neutrality for $\theta=3.15º$ ($i$=10). Rest of parameters like in Fig. \ref{fig:sigma1}(a). Note that $\im \sigma(\omega)$ has been multiplied by $\omega$ to reveal its $\omega\to 0$ asymptotics.}
   \label{fig:sigma2}
\end{figure}

\begin{figure}
   \centering
   \includegraphics[width=0.5\columnwidth]{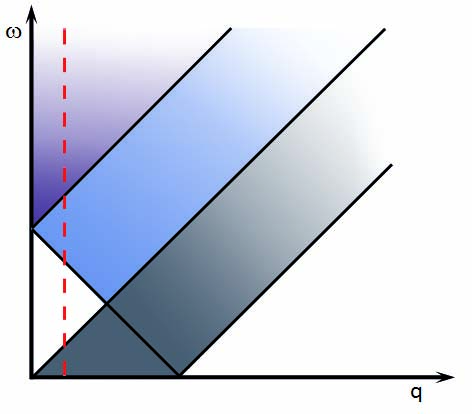} 
      \caption{Two particle spectrum of pristine graphene. Black shaded regions correspond to intraband transitions, whereas blue and violet shaded regions indicate interband transitions. The red vertical line denotes processes with finite momentum $q$. }
   \label{fig:Diagram}
\end{figure}
\begin{figure}
   \centering
   \includegraphics[height=0.4\columnwidth]{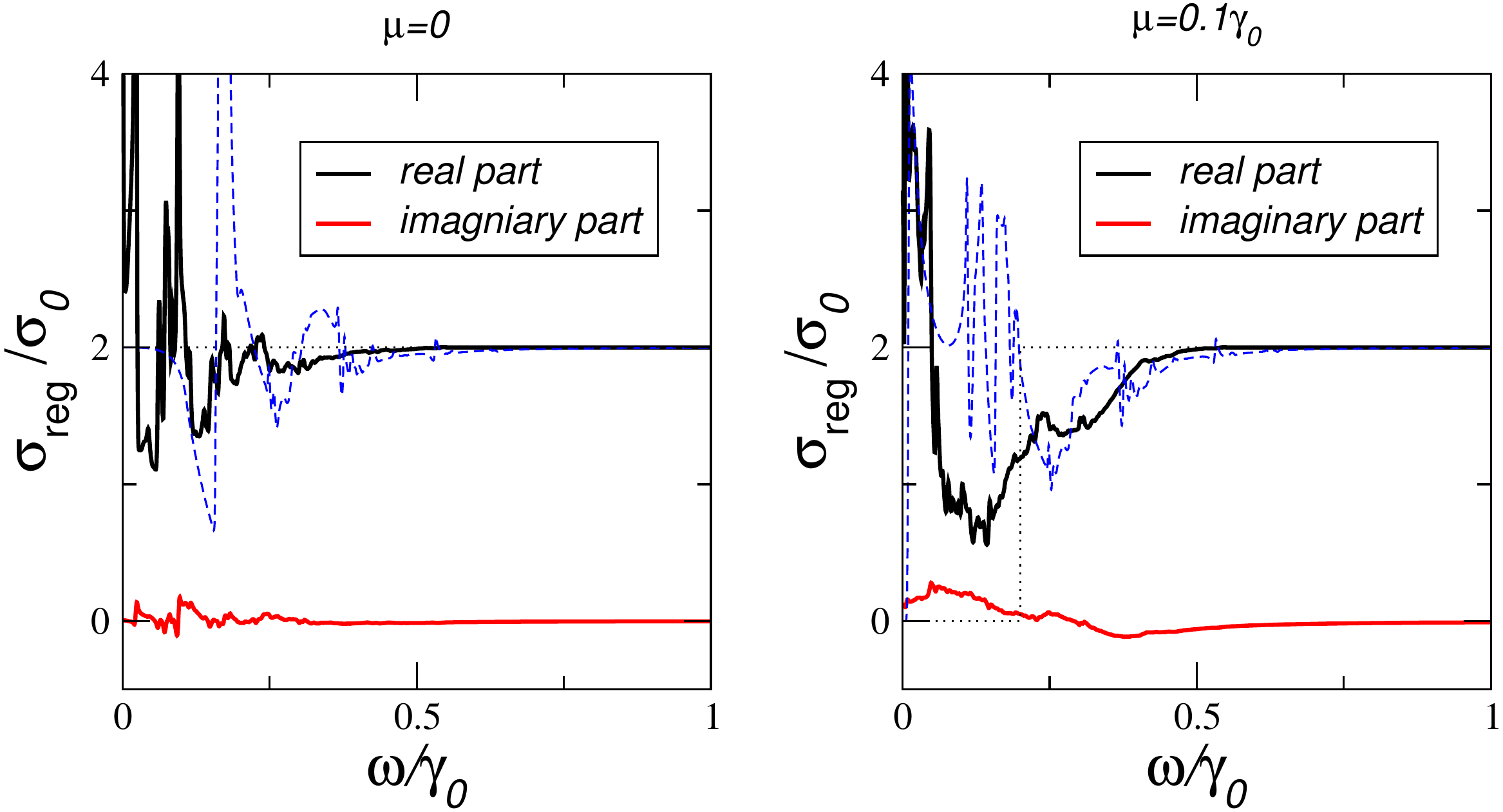}
   \includegraphics[height=0.4\columnwidth]{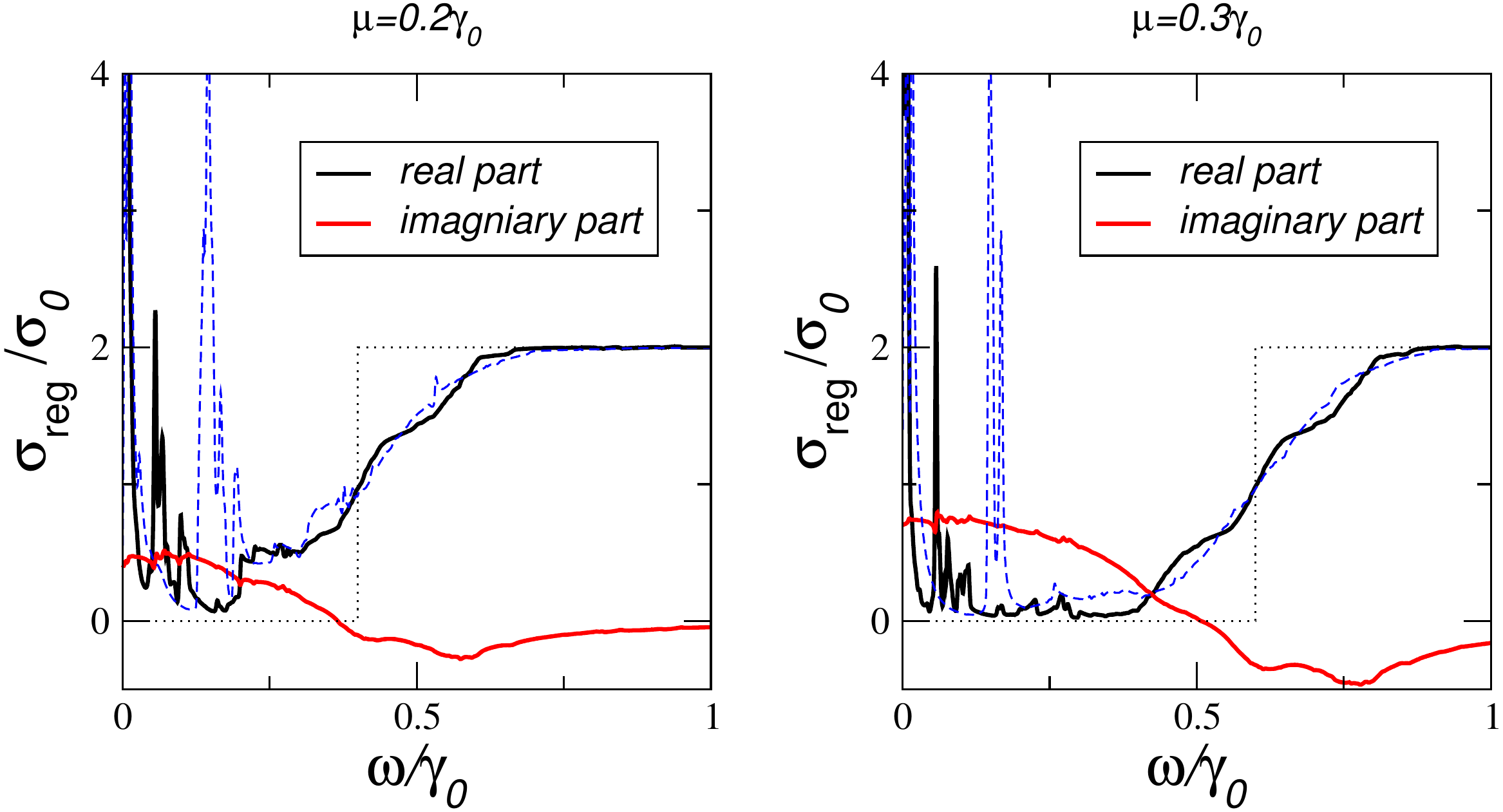}
   \caption{(color online): Real (black) and imaginary (red) part of the optical conductivity for different chemical potential $\mu$ for $\theta_1^m=1.05º$ ($i$=31). Rest of parameters like in Fig. \ref{fig:sigma1}. For comparison, also $\re\sigma$ for ($i=10$) is shown (dashed blue line). Dotted lines resemble the limit for uncoupled layers. Note that $\im \sigma(\omega)$ has been multiplied by $\omega$ to reveal its $\omega\to 0$ asymptotics.}
   \label{fig:sigma31}
\end{figure}
  
The conductivity at neutrality $\mu=0$ is shown in Fig. \ref{fig:sigma1} for two "large" twist angles. Much like the monolayer, the real part (black) exhibits a universal asymptotic value $\re\sigma(\omega\to 0)=2\sigma_0$. It also exhibits signatures of the van-Hove singularities, in form of a dip-peak structure as discussed in detail in Ref. \cite{Moon:13}. The imaginary part (red) exhibits similar features, but vanishes at $\omega\to 0$, i.e., there is no Drude weight at $\mu=0$. This changes for finite temperature where the real part of $\sigma$ is suppressed at $\omega=0$,\cite{Peres08} leading to a finite Drude weight necessary to fulfill the f-sum rule Eq. (\ref{SumRule}). The Drude weight for finite temperature can be deduced from Eq. (\ref{DrudeDirac}) with the substitution $\mu\rightarrow2\ln2k_BT$.\cite{Falkovsky07} Temperature can further provoke a shift of spectral weight close to the van Hove singularity (see the circled region).

Also in the case of finite doping $\mu\neq 0$, see Fig. \ref{fig:sigma2}, the imaginary part of the conductivity develops a $\sim D(\mu)/\omega$ low frequency divergence, while the real part becomes suppressed below a threshold frequency $2\mu$ for small enough $\mu$ (see black and red curves). Unlike for the monolayer, however, this is only true for $\mu$ below the van-Hove singularity, since it relies on the linearity of the bandstructure (and density of states) around $\mu=0$. For chemical potentials close to the van Hove singularity, i.e., $\mu\approx\epsilon_M$, the optical gap is again filled by spectral weight and the Drude weight reduces consistent with the f-sum rule (blue dashed curves). At finite temperature, the optical gap may persist for small  twist angle and the Drude weight can thus even increase, see Fig. \ref{fig:Drude}.

For large chemical potentials $\mu\gg\epsilon_M$, the electronic spectrum (DOS) of the continuum model is just that of two uncoupled graphene layers and one could thus expect to recover the conventional graphene conductivity. But the Moiré coupling between layers induces additional damping, since it opens up the possibility of more interband dissipation channels, i.e., any particle hole excitation with momentum equal to a multiple of the momenta of the Moiré lattice. In order to understand the basic features of the conductivity on the basis of the uncoupled system, we can thus attribute a finite in-plane wavenumber $q\sim n\Delta K$ to the incoming photon ($n>0$), see Fig. \ref{fig:Diagram}. The two-particle spectrum of interband and intraband excitations then predicts dissipation at low energies (black) and a broadened transition (blue) from the optical gap to the universal value $2\sigma_0$ of the conductivity.

In Fig. \ref{fig:sigma31}, the optical conductivity at the first magic angle $i=31$ is shown for various chemical potentials (solid lines) and compared to $\re\sigma_{\rm reg}$ for $i=10$ (dashed blue line). For low chemical potential, the optical response differs considerably for small and large angles, but for increasing $\mu$ the dependence on the twist angle becomes less and $\re\sigma$ is characterized by a large Moire-broadening of the optical gap around $\hbar\omega=2\mu$. Still, there is the pronounced spectral weight at $\hbar\omega=\epsilon_{vH}$ almost independent of $\mu$, which is especially striking for larger angles. This feature cannot be explained from Fig. \ref{fig:Diagram} and is twist angle dependent.

\section{Drude weight}
Considering only intraband transitions of one unbounded conduction band, the Drude weight or charge stiffness is proportional to the chemical potential, independent of its dispersion relation. For parabolic band electrons, $D$ is thus simply related to the charge density, $D=e^2n/m$, whereas for Dirac Fermions one gets $D\sim\sqrt{n}$, see Eq. (\ref{DrudeDirac}). In systems with a finite band-width, we expect a non-monotonous behavior expressing the the fact that the Fermi energy is first related to electrons and later to holes when sweeping the chemical potential through the band. For the one-dimensional tight-binding model, one finds, e.g., $D\sim\sqrt{1-(\mu/2\gamma_0)^2}$ with $2\gamma_0$ the band width. For the two-dimensional hexagonal tight-binding model, the system is characterized by a van Hove singularity at $\mu=\gamma_0$ around which the Drude weight shows a maximum.\cite{Peres08} 

For twisted bilayer graphene described by the continuum model, we expect a combination of both features, because of the unbounded linear dispersion of Dirac Fermions and the superlattice structure of the Moir\'e-pattern. From Fig. \ref{fig:dos} we see that the density of states is characterized by a van Hove singularity followed by a "real" step-like singularity which is repeated to higher doping levels. This pseudo-gap structure divides the spectrum in several bands and will be followed by the Drude weight. Nevertheless, these modulations are superposed onto a constant linear background as suggested by the unbounded continuum model.

\begin{figure}
   \centering
   \includegraphics[height=0.6\columnwidth]{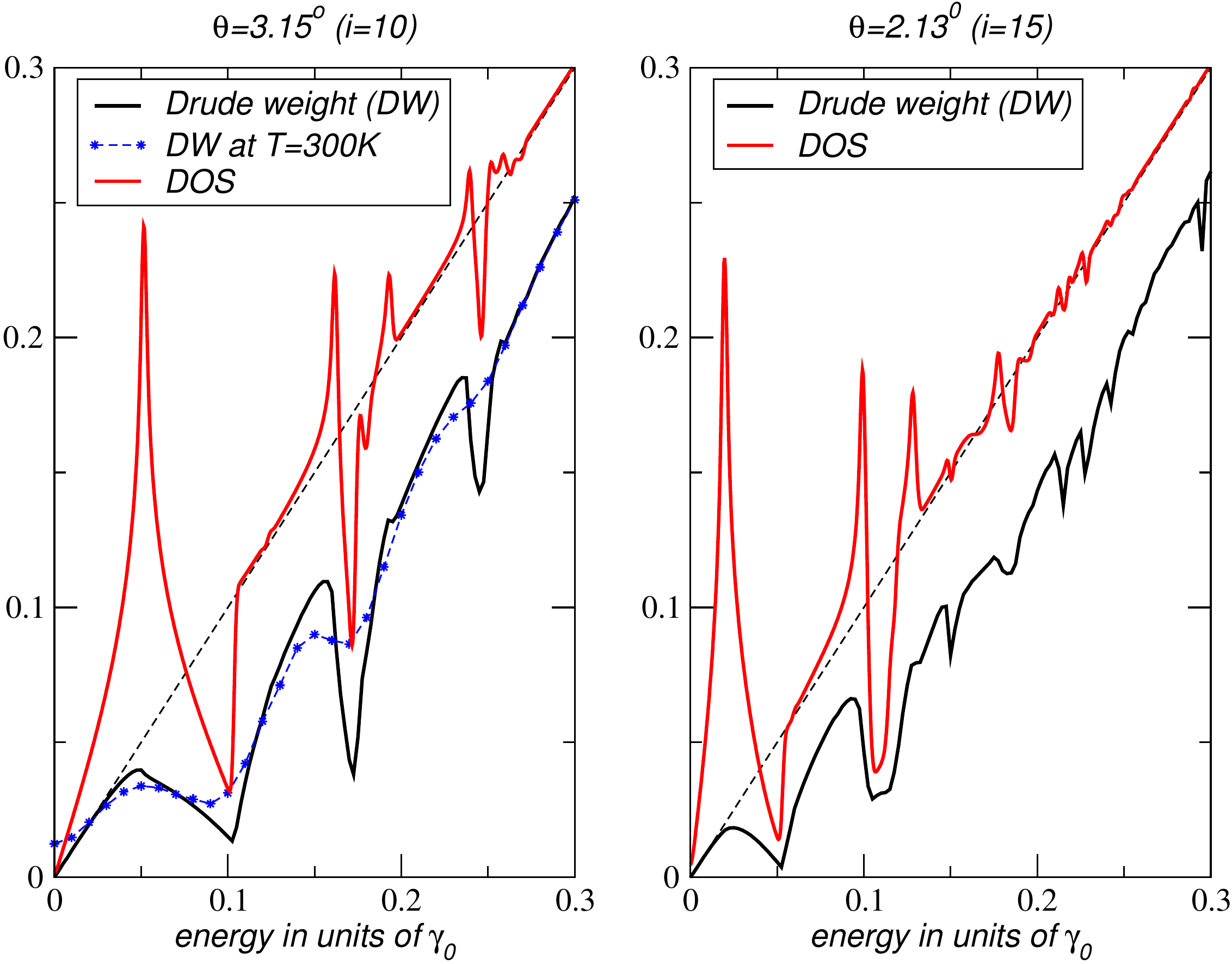}
   \caption{Drude weight $D$ as a function of chemical potential $\mu$ for two different twist angles (black curve), together with the corresponding density of states (DOS) (red curve). The thin dashed line corresponds to the Drude weight and DOS of two decoupled monolayers, in units of $2D_0$, Eq. (\ref{DrudeDirac}), and $\rho_0=\frac{4\gamma_0}{\pi (\hbar v_F)^2}$, respectively. Symbols (blue $\star$) resemble the Drude weight at $T=300$K.}
   \label{fig:Drude}
\end{figure}

The $\mu$ dependence of the Drude weight $D(\mu)$ in units of $2D_0$ is presented in Fig. \ref{fig:Drude} for two large twist angles. As is apparent, for small chemical potential $D(\mu)$ follows twice the monolayer value $2D_0$ independent of the twist angle, as expected for Dirac systems. For larger $\mu$, the Drude weight drops and remains less than $2D_0$ until it converges to the uncoupled bilayer system for $\mu\gg\gamma_0$.
 
 As outlined above, $D(\mu)$ exhibits dips that accurately correlate with dips in the density of states (DOS), also shown in Fig. \ref{fig:Drude} (red). The resulting picture suggests the above mentioned approximate shell structure, albeit (in general) without gaps between shells, just as the suppressed DOS. The first shell can be fairly well approximated be the hexagonal tight-binding model.
 
At finite temperature $T=300$K (and also for disorder), the Drude weight is smeared out for large twist angle $i=10$, thus reaffirming the interpretation of the repeated shell-structure. Nevertheless, the shell structure becomes less pronounced with smaller $\theta$ which is also reflected by its temperature response. 

%Already for $i=15$, the Drude weight at $T=300$K cannot be correlated to its $T=0$K-result, as was already mentioned in Sec. I

\begin{figure}
   \centering
   \includegraphics[height=0.6\columnwidth]{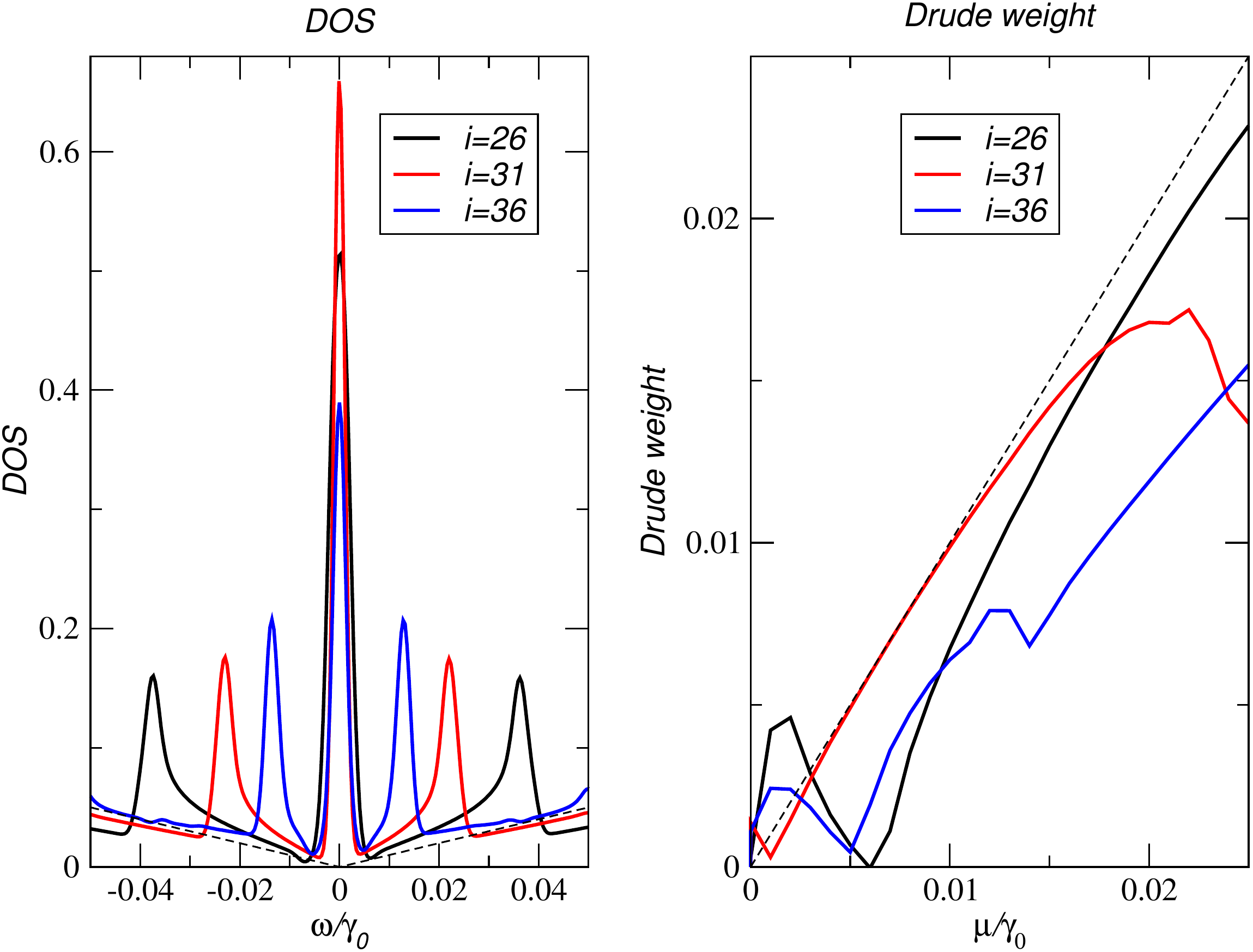}
   \caption{The Density of States (DOS) (left) and the Drude weight (right) around the first critical angle $\theta=1.05º$ ($i=31$) in units of  Fig. \ref{fig:Drude}. Dashed lines correspond to uncoupled bilayer.}
   \label{fig:DrudeDos}
\end{figure}

For small angles, the Drude weight provides a sensitive fingerprint for the breakdown of Dirac physics that occurs around the magical angles due to the emergence of flat bands. This can be seen on the right hand side of Fig. \ref{fig:DrudeDos} for small angles above ($i=26$) and below ($i=36$) the critical angle. The Drude weight becomes {\em larger} than the Drude weight of two monolayers before it drops indicating the band width of the flat band. At the critical angle ($i=31$), this band width becomes zero, and interestingly, the scaling of $D_0$ is recovered. Note that for $i=26$, we find a transport gap $D=0$ at finite chemical potential $\mu=0.06\gamma_0$. 

The increase of the band width for twist angles larger than $\theta_1^m$ can be nicely appreciated from the Drude weight. This becomes considerably more difficult from the DOS, displayed on the left hand side of Fig. \ref{fig:DrudeDos}. The reason for this is that the Drude weight is obtained from integrating over the whole spectral region and not only from the low energy band structure. Nevertheless, the weight of the peak at zero energy is maximal at the magic angle (red curve), indicating its singularity.


\begin{figure*}
   \centering
   \includegraphics[height=0.4\columnwidth]{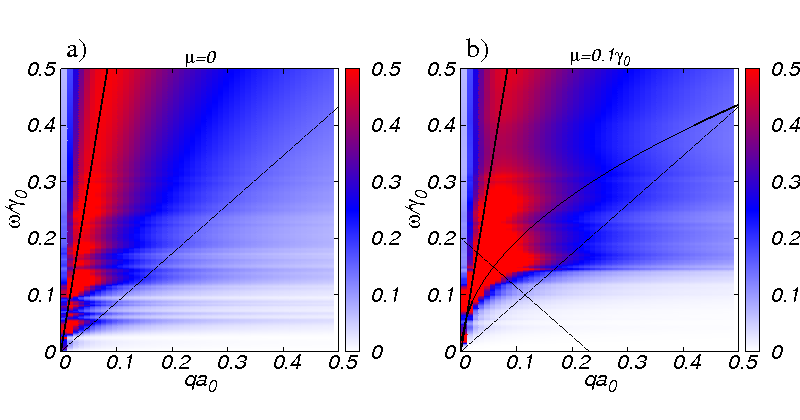} 
   \includegraphics[height=0.4\columnwidth]{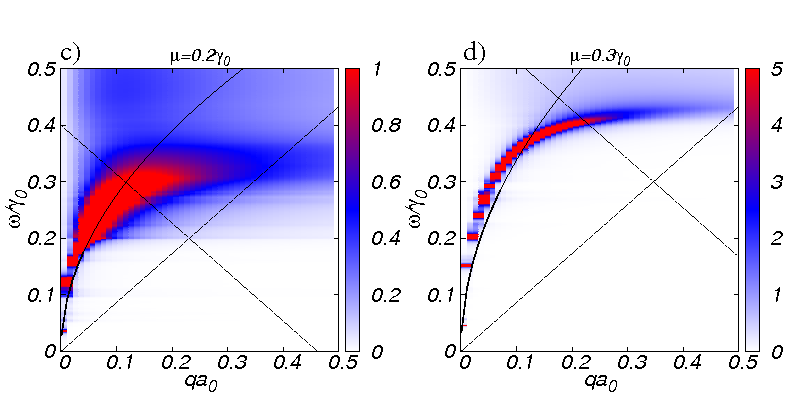} 
      \caption{Loss function $S(\bm{q},\omega)=-\im \epsilon^{-1}(\bm{q},\omega)$ in the long-wavelength RPA approximation. The dashed curved line corresponds to the undamped plasmon of the decoupled bilayer. The region of undamped plasmons of the decoupled bilayer is defined by straight lines.}
   \label{fig:plasmons}
\end{figure*}

\section{Plasmons}
A complementary aspect to the optical absorption is the plasmon excitation spectrum of the electron gas. Conceptually, a plasmon is an oscillating charge density mode (hence accompanied by an oscillating electric potential), which is sustained by the Coulomb interactions between electrons. The dispersion relation of these modes is given by the zeroes of the dielectric function $\epsilon(\bm{q},\omega)=0$, that relates the screened electric potential $V_\mathrm{sc}(\bm{r},t)$ to an externally applied potential $V_\mathrm{ext}(\bm{r},t)$,
\[
V_{sc}(\bm{q},\omega)=\frac{V_\mathrm{ext}(\bm{q},\omega)}{\epsilon(\bm{q},\omega)}\;.
\]
A plasmon is, therefore, a finite solution $V_{sc}(\bm{q},\omega)$ that requires no external driving, i.e., it is self-sustained. A Landau-damped plasmon corresponds to a \emph{peak} in $1/\epsilon(\bm{q},\omega)$, and decays due to dissipation into particle-hole excitations.

The dielectric function $\epsilon(\bm{q},\omega)$ may be computed within the Random Phase Approximation (RPA), which in two dimensions yields
\[
\epsilon_\mathrm{RPA}(\bm{q},\omega)=1-\frac{e^2}{2\varepsilon_0\epsilon |\bm{q}|}\chi_{\rho\rho}(\bm{q},\omega)
\]
where $\chi_{\rho\rho}$ is the charge susceptibility, and $\epsilon$ is the dielectric constant of the environment surrounding the two-dimensional electron gas. From the continuity equation between charge and current, it follows that \cite{Giuliani:05}
\[
\chi_{\rho\rho}(\bm{q},\omega)=\frac{q^2}{\omega^2}\chi_{jj}^L(\bm{q},\omega)
\]
where $\chi_{jj}^L$ is the longitudinal part of the current-current correlator in Eq. (\ref{Kubo}). In the long-wavelength limit, we may thus relate $\epsilon_\mathrm{RPA}$ directly to $\sigma(\omega)$ using Eq. (\ref{sigma}),
\begin{equation}
\label{eRPA}
\epsilon_\mathrm{RPA}(\bm{q},\omega)\approx 1+\frac{i}{2\varepsilon_0\epsilon}\frac{|\bm{q}|}{\omega}\sigma(\omega)=1+i\frac{\pi\alpha_g}{2\epsilon} \frac{v_F |\bm{q}|}{\omega}\frac{\sigma(\omega)}{\sigma_0}.
\nonumber
\end{equation}
where $\alpha_g=e^2/(4\pi\epsilon_0 \hbar v_F)\approx 2.2$ is graphene's fine structure constant.
 
To compute $\omega_p$ for small frequencies, we may expand $\im \sigma$ in Eq. (\ref{Imsigma}) by the Drude term,
\[
\im \sigma(\omega)\approx \frac{D}{\omega}\;.
\]
%where $D'$ is defined as
%\[
%D'=\lim_{\omega\to 0}\mathcal{P}\int_0^\infty d\nu \frac{\re\sigma_{xx}^\mathrm{reg}(\nu)-2\sigma_0}{\nu^2-\omega^2}\approx \frac{2\sigma_0}{\omega_t}
%\]
%(Note that for two decoupled monolayers, the above expression holds exactly, with $\omega_t=2\mu/\hbar$). 
This results in the dispersion of a classical 2D plasmon
\begin{equation}
\label{PlasmonDispersion}
\omega_p\approx \sqrt{\frac{D|\bm{q}|}{2\varepsilon_0\epsilon}}\;.
\end{equation}
For two decoupled monolayers, one just replaces $D=8\mu/\pi\hbar\sigma_0$ above, which is shown as dotted curved line in Fig. \ref{fig:plasmons}.

\subsection{Plasmons in twisted Bilayer}
In twisted graphene bilayer there may exist four types of plasmonic modes or resonances: i) undamped (conventional) graphene plasmons for chemical potentials with $\mu\ll\epsilon_M$, ii) interband "plasmons" present also for $\mu=0$, iii) damped plasmons for large chemical potential $\mu\gg\epsilon_M$, iv) transverse plasmons. Below, we will briefly outline the main properties of these different types:
 
\subsubsection{Undamped plasmons}
At low doping, there exists a low energy window with $\re\sigma=0$ that gives rise to undamped plasmons governed by the Drude weight $D$. In this regime, the twisted bilayer can be described by linear Dirac Fermions and $D(\mu)$ follows the Drude weight of the uncoupled system $2D_0$, as shown in Fig. \ref{fig:Drude}. This energy window becomes smaller for decreasing twist angle since the van Hove singularity moves closer to the neutrality point and is only relevant for large angles. 

\subsubsection{Interband plasmons}
In monolayer graphene, a peak in the loss function associated to ${\bm\pi}\rightarrow{\bm\pi}^*$ transitions around the van Hove singularity and with linear dispersion was first predicted by DFT-studies,\cite{Kramberger08} and later experimentally observed in suspended graphene by electron energy-loss spectroscopy (EELS).\cite{Eberlein08} Within the hexagonal tight-binding model and RPA, i.e., without including correlation or renormalization effects, no zero of the dielectric function is obtained.\cite{Stauber10a} The absorption peak would thus be merely due to interband transitions enhanced by a band-structure effect; nevertheless, in bi- or multilayer, $\epsilon^{RPA}(\bm{q},\omega)$ becomes zero around the M-point.\cite{Yuan11} In the twisted graphene bilayer, there exist several van Hove singularities such that interband plasmons (or absorption enhancement) are to be expected.

\subsubsection{Damped plasmons}
In the regime of large chemical potential, $\mu\gg\epsilon_M$, the conventional intraband plasmon is recovered, albeit Moire-damped. This can be deduced from Fig. \ref{fig:sigma31} which will be used to discuss the plasmon dispersion in terms of the loss function. The dispersion only depends slightly on the twist angle and becomes well-defined for large $\mu$, extending into the Landau damped region just as for the monolayer. 

\subsubsection{Transverse Plasmons}
There is also the possibility for transverse or transverse electromagnetic (TE) plasmons modes given by the condition\cite{Ziegler07}
\begin{equation}
1-\frac{\omega i \sigma(\omega)}{c^2\sqrt{q^2-\omega^2/c^2}}=0\;.
\label{TE}
\end{equation}
For TE modes, the plasmon dispersion is closely pinned to the light cone which justifies our local approximation ($q\to 0$) and there are (damped) solutions to Eq. (\ref{TE}) if $\im\sigma<0$. Since they can only exist in twisted bilayer surrounded by a homogeneous dielectric background,\cite{Kotov13} we will not discuss them any further.

\subsection{Loss Function} 

The loss function, defined as $S(\bm{q},\omega)=-\im \epsilon^{-1}(\bm{q},\omega)$ is a measure of the spectral density of plasmon excitations. Thus, a sharp peak in $S(\bm{q},\omega)$ reveals a long-lived (transverse magnetic, TM) plasmon of that frequency and momentum. A proper undamped plasmon, defined by $\epsilon(\bm{q},\omega)=0$, corresponds to a delta peak in $S(\bm{q},\omega)$, since $\epsilon^{-1}(\bm{q},\omega)$ satisfies Kramers-Kronig relations. 

In Fig. \ref{fig:plasmons}, we present $S(\bm{q},\omega)$ within the long-wavelength RPA approximation of Eq. (\ref{eRPA}) at the first magic angle $\theta_1^m=1.05º$ ($i=31$) for different chemical potential with $\epsilon=1$. Similar results are obtained for larger angles, as can be deduced from Fig. (\ref{fig:sigma31}). In all cases, the Dirac dispersion $\omega=v_Fq$ and the $\hbar\omega=2E_F-\hbar v_Fq$ from Fig. \ref{fig:Diagram} is shown as dashed lines, indicating the onset of intra- and interband transitions. For zero or low chemical potential (Fig. \ref{fig:plasmons}a), the loss function indicates an acoustic interband "plasmon" with sound velocity $v_s\approx\pi\alpha_gv_F$ (solid line),\footnote{It is not a plasmon in the conventional sense since the dielectric function does not become zero.} and only for low energies signatures of the van Hove singularities are seen which are more pronounced for larger angles. This linear (interband) mode crosses over to the conventional $\sqrt{q}$ (intraband) mode, Fig. \ref{fig:plasmons}b) and c), and becomes well-defined for large chemical potential, Fig. \ref{fig:plasmons}d, although there are slight variation compared to Eq. (\ref{PlasmonDispersion}) shown as solid curved line.

\section{Conclusions}

We have analyzed the density of states (DOS), optical conductivity, Drude weight, and plasmon spectrum in a twisted bilayer for various angles, doping levels and temperature, and compared them to the uncoupled system. We have found a universal low-frequency conductivity and peaks at finite frequency that correlate to van-Hove singularities in the DOS. These latter features, however, become scrambled at decreasing angles. We have also shown that the Drude weight in twisted bilayers is usually below that of two decoupled monolayers, $2D_0$, and exhibits a shell-like structure, following again van-Hove singularities in the DOS. Only for small angles close to the neutrality point, the Drude weight becomes larger than $2D_0$, signaling the emergence of a flat band. The Drude weight resembles a sensitive quantity to numerically distinguish the different electronic structure at critical angles since it is obtained from integrating over the whole spectrum and is directly related to transport experiments. We have finally discussed how plasmons may become damped by the interlayer Moir\'e coupling which opens up new dissipation channels. Significant spectral structure survives, however, including undamped plasmons and multiple long-lifetime resonances (interband plasmons) within the extended interband particle-hole continuum.

\section{Acknowledgments} We thank G. G\'omez-Santos and E. Prada for discussions. This work has been supported by FCT under grants PTDC/FIS/101434/2008; PTDC/FIS/113199/2009, by MIC-Spain under grant FIS2010-21883-C02-02, FIS2011-23713 and FIS2012-33521, and by the European Research Council Advanced Grant, contract 290846.

\section*{References}

\bibliographystyle{unsrt}

%\bibliography{twisted}

\end{document}